\documentclass{aa}
\usepackage{graphicx}
\begin{document}

\title{Early-type variables in the Magellanic Clouds}
\subtitle{I. $\beta$~Cephei stars in the LMC bar}

\author{A.~Pigulski, Z.~Ko{\l }aczkowski}

\offprints{A.~Pigulski}

\institute{Wroc{\l }aw University Observatory,
          Kopernika 11, 51-622 Wroc{\l }aw, Poland}

\date{Received 13 February 2002 / Accepted 21 March 2002}

\abstract{A thorough analysis of the OGLE-II time-series photometry of
the Large Magellanic Cloud bar supplemented by similar data from 
the MACHO database led us to the discovery of three $\beta$ Cephei-type
stars.  These are the first known extragalactic $\beta$~Cephei-type
stars.  Two of the three stars are multiperiodic.  Two stars have
inferred masses of about 10~$M_\odot$ while the third is about 2~mag
brighter and at least twice as massive.  All three variables are
located in or very close to the massive and young LMC associations
(LH\,41, 59 and 81).  It is therefore very probable that the variables
have higher than average metallicities.  This would reconcile our
finding with theoretical predictions of the shape and location of the
$\beta$~Cephei instability strip in the H-R diagram.  The low number
of $\beta$~Cephei stars found in the LMC is another observational
confirmation of strong dependence of the mechanism driving pulsations
in these variables on metallicity.  Follow-up spectroscopic
determination of the metallicities in the discovered variables will
provide a good test for the theory of pulsational stability in
massive main-sequence stars.
}
\titlerunning{$\beta$~Cephei stars in the LMC bar}
\maketitle

\section{Introduction}
Since the early 90-ties it is known that pulsations in $\beta$~Cephei
stars are driven by $\kappa$ mechanism operating 
at a temperature around 2$\times$10$^5$ K where the opacity bump occurs 
(Cox et al.~\cite{cox92};
Moskalik \& Dziembowski \cite{modz92}; Kiriakidis et
al.~\cite{kiri92}; Dziembowski \& Pamyatnykh \cite{dzpa93}; Gautschy
\& Saio \cite{gasa93}).  The bump is caused by a large number of
bound-bound transitions in the iron-group ions.  As a consequence, the
area of the instability strip in the Hertzsprung-Russell (H-R) diagram 
should strongly depend on metallicity (Moskalik \& Dziembowski 
\cite{modz92}; Pamyatnykh \cite{pamy99}).  The instability in models
of $\beta$~Cephei stars results from a small domination of driving
over damping (see, e.g., Moskalik \& Dziembowski \cite{modz92}), thus the
subtle differences in both the input physics and the modeling may lead to
considerable differences in the predicted position of the instability strip.  For
instance, for solar metallicity, Pamyatnykh (\cite{pamy99}) finds that the
$\beta$ Cephei instability strip (hereafter BCIS) has no upper edge
for the most massive stars and extends beyond the core-hydrogen
burning phase of evolution.  Moreover, for the heavy elements
abundance parameter $Z$ = 0.01, his BCIS shrinks down to a small
region at high luminosities, where $\beta$ Cephei stars have not been found.
On the other hand, for solar abundance, Deng \& Xiong
(\cite{dexi01}) predict the much narrower BCIS with an upper boundary for
$M \approx$ 20~$M_\odot$.  In their calculations the BCIS disappears at
$Z$ as low as 0.005.

In this context, the observational study of $\beta$ Cephei star
pulsations in objects of different metallicity is of great importance.
It was already pointed out by Sterken \& Jerzykiewicz (\cite{stje88})
that with their lower than Galactic metallicities and relatively small
interstellar absorption, the Large and Small Magellanic Clouds (LMC
and SMC) are among the best objects for such a study.  Although
typical photometric amplitudes of $\beta$ Cephei stars are of the order of 0.01~mag
and these stars should not be brighter than $\sim$14~mag in the LMC and
even fainter in the SMC, with the modern techniques allowing the CCD
photometry in crowded fields, the detection of these stars is
certainly within reach.

Searches for $\beta$ Cephei stars in the Magellanic Clouds had already
been undertaken.  The first was carried out by Sterken \&
Jerzykiewicz (\cite{stje88}).  By means of the photoelectric
photometry these authors studied six late O/early B-type stars.  For
the same purpose, Kubiak (\cite{kubi90}) searched the young LMC
cluster NGC\,1712.  These searches resulted in the discovery of some
variables, but none of a convincing $\beta$ Cephei-type pulsation 
mainly because of the small statistical sample observed.  Another search 
for $\beta$ Cephei stars was
performed by Balona (\cite{balo92}; \cite{balo93}) and Balona \&
Jerzykiewicz (\cite{baje93}) in NGC\,2004 and 2100 in the LMC and in
NGC\,330 in the SMC.  A similar CCD search was also carried out by 
Kjeldsen \& Baade (\cite{kjba94}) in NGC\,2122 in the LMC and NGC\,371 in
the SMC. No variable of $\beta$~Cephei-type
was found by these authors.  

The reduction of the OGLE-II data (Udalski et al.~\cite{udal97}) for
the Magellanic Clouds by means of the Difference Image Analysis
(hereafter DIA, Wo\'zniak \cite{wozn00}; \.Zebru\'n et
al.~\cite{zebr01a}) yielded the photometry for over 53,000 candidate
variables in the LMC and 15,000 in the SMC.  A catalog including these stars
was recently made available to the astronomical community (\.Zebru\'n
et al.~\cite{zebr01b}).  Earlier, the OGLE $BVI_{\rm C}$
photometry of about 6.7 $\times$ 10$^6$ stars in the LMC (Udalski et
al.~\cite{udal00}) and 2.2 $\times$ 10$^6$ stars in the SMC (Udalski et
al.~\cite{udal98}) was published.  At the same time, the even larger 
database of the MACHO observations of 7.3 $\times$ 10$^7$ stars in both Clouds and 
in two filters ($V$, $R$) has become available (Allsman \& Axelrod \cite{alax01}).

These databases offer an unprecedented opportunity to search for the
presence of $\beta$ Cephei-type stars in the LMC and SMC.  Combined with
the detailed metallicity determinations, this result will put strong
constraints on the theoretical stability predictions.  As we shall show
later, the accuracy of the photometry is good enough to detect all
$\beta$ Cephei-type pulsations in the LMC stars with semi-amplitudes exceeding
3 to 5~mmag, depending on the star brightness.

In this paper, intended to be the first of a series, we present
the results of the search for $\beta$~Cephei stars among the LMC bar stars
using the OGLE-II DIA photometry.  A similar analysis for the SMC stars will be
published separately.  The details of the analysis and the selection
of stars is described in Sect.~2.  Section 3 presents the main results
of the search.  The possible connection of the variables we found with
the LMC clusters and associations is verified in Sect.~4.  We also
briefly discuss the variables detected during previous attempts to
find $\beta$~Cephei stars in the LMC (Sect.~5).  The consequences of our
finding are discussed in detail in Sect.~6, while a short summary and
our plans concerning the next papers of the series are given in the
last section.

\section{Data selection and analysis}
The output of the DIA is the photometry of the variable stars only
(strictly speaking the candidates for variable stars, as the DIA
produces also artifacts), therefore it was an obvious choice to take
the OGLE-II catalog of about 53,400 candidate variables in the LMC
(\.Zebru\'n et al.~\cite{zebr01b}) as the main source of the
time-series photometric data.  For the details of the OGLE-II
observations and the DIA reductions we direct the reader to the paper of
\.Zebru\'n et al.~(\cite{zebr01a}).  We only mention that the OGLE-II
observations of the LMC cover twenty-one 14$\farcm$2 $\times$ 56$\farcm$9
fields in the bar of this galaxy.  For a single star, about 30, 40,
and 400 observations are typically available in the $B$, $V$ and $I_{\rm
C}$ bands, respectively.  The OGLE-II observations span at most
3.5~yr, from the beginning of 1997 to mid-2000.

%
%
\begin{figure}
\resizebox{\hsize}{!}{\includegraphics{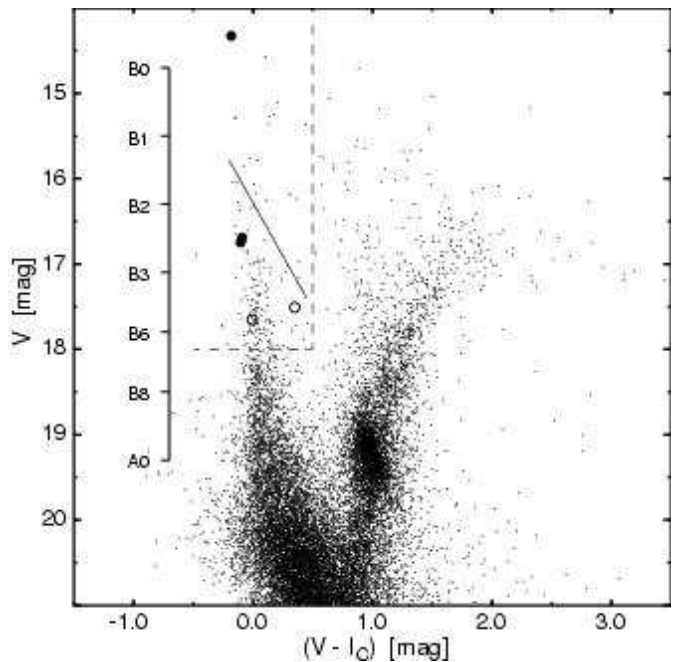}}
\caption{Colour-magnitude diagram for $\sim$16,000 stars in the
OGLE-II LMC field SC1, randomly selected from the photometry of
Udalski et al.~(\cite{udal00}).  The three $\beta$ Cephei stars, V1,
V2, and V3, are shown as filled circles, the other two short-period
variables, V4 and V5, as open circles.  For reference, the location of
the luminosity class V B-type stars is shown considering an LMC distance modulus
of 18.5~mag and $A_{\rm V}$ = 0.4~mag.  The inclined solid line is the
reddening line for $A_{\rm V}/E(V-I_{\rm C})$ = 2.49 (Stanek
\cite{stan96}).  The dashed line shows the limits of the search for
$\beta$~Cephei variables.}
\label{f_cmd}
\end{figure}

The MACHO data archive (Allsman \& Axelrod \cite{alax01}) served as
the second source for the time-series data.  The data span about 8
years between mid-1992 and the beginning of 2000, and are available in
two passbands:  blue (440--590~nm) and red (590--780~nm).  Henceforth
we shall refer to these passbands as $V_{\rm M}$ and $R_{\rm M}$,
respectively.  The MACHO observations cover a much larger area around
the LMC than the OGLE-II ones do.  However, the database is at present
accessible in an unsuitable way for our search.
We therefore use here the MACHO archive only for stars we found particularly
interesting from the analysis of the OGLE-II data and for variables
detected in previous searches (see Sect.5).  A thorough search for
$\beta$ Cephei stars in the MACHO archive will be done as soon as
enhanced search capabilities become available.

Prior to the analysis, the observations with larger photometric errors have been
rejected from the MACHO data.  In addition, some outliers were
removed.  Finally, the heliocentric corrections were applied to the
MACHO epochs.  Since the published epochs correspond to the beginning of an
exposure (Cook \cite{cook02}), 150 seconds correction (half the exposure
time) was added to them.  The OGLE-II epochs have been corrected for the 
'drift-scan effect' according to the equation given by \.Zebru\'n et al.~(\cite{zebr01b}).

The coldest and faintest Galactic $\beta$~Cephei stars have MK
type of about B2.5\,V, corresponding to $M_{\rm V}$ $\approx$ $-$2.
Taking into account the LMC distance modulus of 18.5~mag, we get $V$ =
16.5~mag.  The average $E(B-V)$ colour excess for the LMC is about
0.13~mag (Massey et al.~\cite{mass95}), yielding $A_{\rm V} \approx$
0.4~mag.  However, the visual extinction is sometimes much larger,
especially for hot stars (Zaritsky \cite{zari99}).  This is obvious,
because there is usually a large amount of interstellar matter around
young hot objects.  In order to take that into account, we decided to add
an 1~mag margin to the limiting magnitude of our search.  Considering also 
the effect of reddening on the photometric indices, we have finally chosen stars with $V <$
18~mag and $(V-I_{\rm C}) <$ 0.5~mag (see Fig.~\ref{f_cmd}) to search
for the presence of $\beta$~Cephei-type
variability.  A total of 5204 stars in the LMC were extracted from the catalogue of
\.Zebru\'n et al.~(\cite{zebr01b}).

%
\begin{table*}
\caption{The $\beta$~Cephei stars in the LMC bar}
\begin{tabular}{ccccccc}
\hline\noalign{\smallskip}
  & OGLE-II & OGLE-II & MACHO & $V$ & $B - V$ & $V - I_{\rm C}$
\\ Name & field & name & name & [mag] & [mag] & [mag] \\
\noalign{\smallskip}\hline\noalign{\smallskip}
V1 & LMC-SC1 & OGLE053446.82$-$694209.8 & 81.8881.161 & 16.691 &
$-$0.109 & $-$0.089 \\
V2 & LMC-SC3 & OGLE052809.21$-$694432.1 & 77.7792.493 & 16.748 &
+0.003 & $-$0.103 \\
V3 & LMC-SC7 & OGLE051841.98$-$691051.9 & --- & 14.327 &
$-$0.175 & $-$0.181 \\
\noalign{\smallskip}\hline
\end{tabular}
\end{table*}

For all these stars we performed frequency analysis by means of the
AoV periodogram of Schwarzenberg-Czerny (\cite{schw96}) in the range
between 0 and 20~d$^{-1}$.  The frequencies of the highest peaks in
all periodograms were then sorted and the photometry of stars with the
highest frequencies was examined visually.

In the classical approach, $\beta$~Cephei stars are recognized as
early B-type stars with periods shorter than 0.3~d (Sterken \&
Jerzykiewicz \cite{stje93}).  Since it seems that $\beta$~Cephei stars
with periods slightly longer than 0.3~d exist (Krzesi\'nski \&
Pigulski \cite{krpi97}), we decided to consider all stars showing
periods shorter than 0.35~d as potential variables of this type.

As a result of our periodogram analysis, we found three stars which we
regard to be the LMC $\beta$ Cephei stars.  Two other short-period
objects were found among early-type variables.  They could also be
$\beta$~Cephei stars, but because another interpretation of their
variability is more likely, we describe them only briefly in
Sect.~3.4.

\begin{figure}
\resizebox{\hsize}{!}{\includegraphics{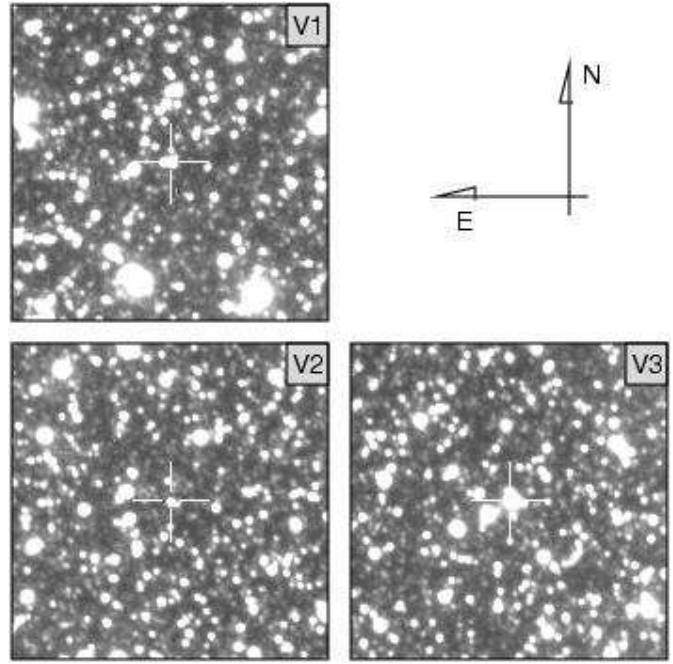}}
\caption{OGLE-II maps (1$\farcm$17 $\times$ 1$\farcm$17) for the three
$\beta$~Cephei stars (V1--V3) in the LMC bar.  The J2000 coordinates
of the stars are coded within the OGLE-II names (Table 1).}
\label{f_findc}
\end{figure}

For the sake of simplicity we designate the $\beta$~Cephei stars as V1
to V3 and list them in Table 1.  The cross-identifications with
the OGLE-II and MACHO designations are also given.  The
finding charts shown in Fig.~\ref{f_findc} were taken from the OGLE-II
database.
%
%

\section{$\beta$ Cephei stars}

\subsection{V1 = OGLE\,053446.82$-$694209.8 = MACHO\,81.8881.161}
\begin{figure}
\resizebox{\hsize}{!}{\includegraphics{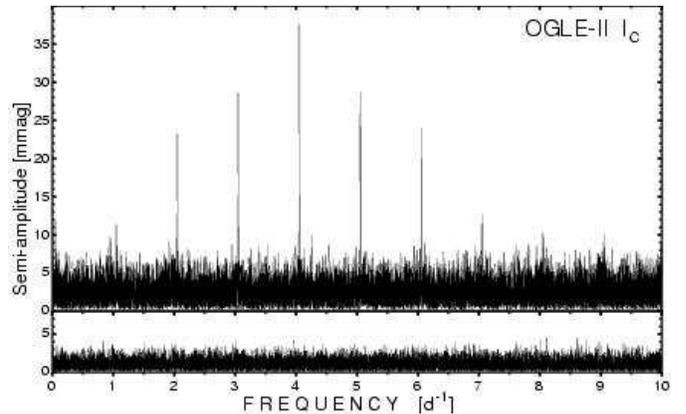}}
\caption{Fourier periodograms of the OGLE-II 1997--2000 $I_{\rm
C}$-filter observations of V1.  The upper panel shows the periodogram
of the original data; the lower one, after prewhitening with $f$ =
4.05160~d$^{-1}$.}
\label{f_p_v1}
\end{figure}

The star is located in the field SC1 and 81 of the OGLE-II and MACHO,
respectively.  Fourier analysis of the OGLE-II $I_{\rm C}$
observations of this star (Fig.~\ref{f_p_v1}) revealed a single
frequency $f$ = 4.05160~d$^{-1}$.  No other periodicity above the
noise level ($\sim$5 mmag) has been found.  Practically the same
result, that is, a monoperiodic signal, has been detected in the MACHO
data except that after prewhitening with $f$, a small peak appeared at
frequency of 1~(sidereal day)$^{-1}$.  It is presumably of
instrumental origin.  The star has virtually the same amplitudes in
the $V_{\rm M}$ and $R_{\rm M}$ bands (see Table 2).  Since the MACHO
data span a wider time interval than the OGLE-II ones do, the refined
frequency, $f$ = 4.051593~d$^{-1}$, has been obtained by means of a
non-linear least squares fitting of the MACHO data.  The semi-amplitudes and phases
of both datasets are given in Table 2.

%

\subsection{V2 = OGLE\,052809.21$-$694432.1 = MACHO\,77.7792.493}
\begin{figure}
\resizebox{\hsize}{!}{\includegraphics{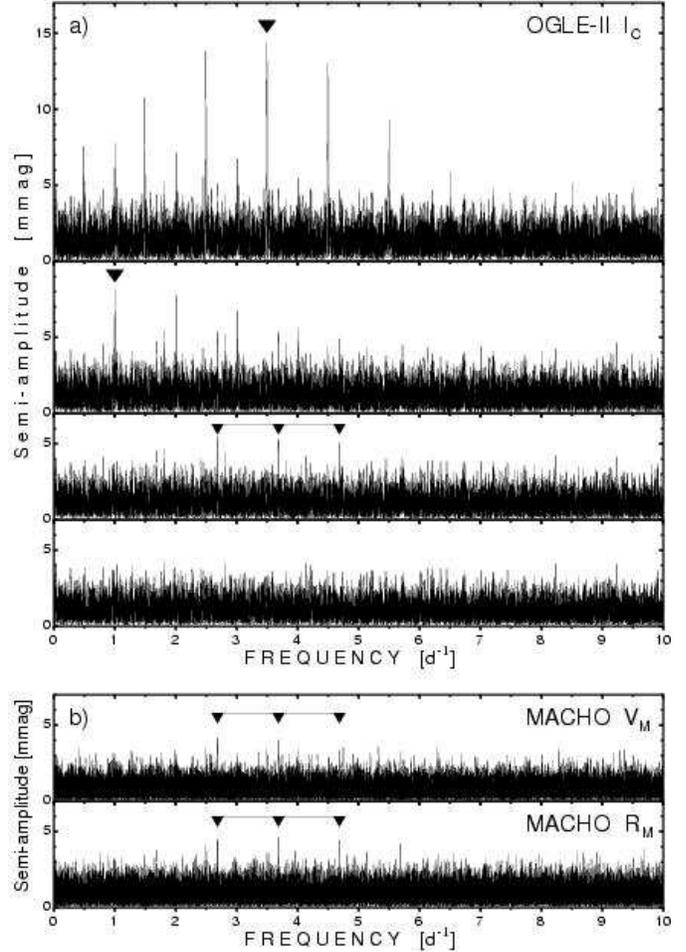}}
\caption{{\bf a.} Fourier periodograms of the OGLE-II 1997--2000
$I_{\rm C}$-filter observations of V2.  From top to bottom the
panels show the periodograms (i) of the original data, (ii) after
prewhitening with $f_1$ = 3.49767~d$^{-1}$, (iii) after prewhitening
with $f_1$ and $f_{\rm sd}$ = 1.00333~d$^{-1}$, (iv) after removing
$f_1$, $f_{\rm sd}$, and $f_2$ = 3.68377~d$^{-1}$.  {\bf b.} Fourier
periodograms of the MACHO $V_{\rm M}$ (top) and $R_{\rm M}$ (bottom) data
after prewhitening with the low-frequency variations and $f_1$.  The
triangles indicate the location of $f_2$ and its 1~d$^{-1}$ aliases.}
\label{f_p_v2}
\end{figure}

The second $\beta$ Cephei-type variable has been found in the OGLE-II
field SC3 and in the MACHO field 77.  It has almost the same colour and
magnitude as V1 (see Fig.~\ref{f_cmd} and Table 1).  The Fourier
periodogram of the OGLE-II $I_{\rm C}$-filter data
(Fig.~\ref{f_p_v2}a) shows the strongest signal at frequency $f_1$ =
3.49767~d$^{-1}$.  After removing this signal the highest peak appears
at a frequency $f_{\rm sd}$ = 1.00333~d$^{-1}$, which is close to
1~(sidereal day)$^{-1}$.  Usually, such a signal appears in the
periodogram when long-term irregular changes are superimposed.  Since
it appears for V2 in all three datasets, it can be caused
by a low-frequency intrinsic variation.  In fact, changes in the
mean magnitude of this star are well seen in the MACHO data:  during
the first three years of observations the mean brightness of V2 faded
by about 0.02~mag.   After
removing $f_1$ and the low-frequency signals, a second periodicity,
$f_2$ = 3.68377~d$^{-1}$, with a very small amplitude appears slightly
above the noise level (Fig.~\ref{f_p_v2}a).  In order to make sure 
this is a real frequency, we
calculated similar periodograms for the MACHO data (Fig.~\ref{f_p_v2}b).
Although barely visible in the $V_{\rm M}$
observations, $f_2$ appears clearly in the $R_{\rm M}$ data.  We
conclude that the frequency is real, and thus V2 is a biperiodic
variable.  However, because of the low amplitude of this periodic
signal, a few close peaks of similar height occur around $f_2$. 
The highest peak does
not have the same frequency in all periodograms; it is equal to
3.6838, 3.6853, and 3.6847~d$^{-1}$ for the OGLE-II $I_{\rm C}$, MACHO
$V_{\rm M}$ and MACHO $R_{\rm M}$ data, respectively.  In the final
fit we used the frequency corresponding to the highest detected peak
among the different
periodograms (see Table 2), but we cannot be certain that our choice is
correct.

%
%

%
%
\begin{table*}
\caption[]{The results of the sine curve fits to the photometric data
of V1--V3.  $N_{\rm obs}$ is the number of individual observations and
RSD is the residual standard deviation.  See Sect.~3.5 for the
definition of the signal-to-noise (S/N) ratio.}
\begin{tabular}{cccccrrccr}
\hline\noalign{\smallskip}
    & Frequency & Period & Data & & &Semi-amplitude& HJD
of the time & RSD & \\
Star& [d$^{-1}$] & [d] & source & Filter& $N_{\rm obs}$&
[mmag] & of maximum light$^{\rm a}$ & [mmag] & S/N\\
\noalign{\smallskip}\hline\noalign{\smallskip}
V1 & 4.051593 & 0.2468165 & OGLE-II& $B$ & 22 & 45.1 $\pm$ 4.8 &
51\,137.1511 $\pm$ .0043 & 15.5 & 7.8\\
& & & OGLE-II& $V$ & 45 & 37.7 $\pm$ 3.1 & 51\,058.4247 $\pm$ .0036 &
15.5 & 9.3\\
& & & OGLE-II& $I_{\rm C}$ & 350 & 38.9 $\pm$ 1.0 & 51\,027.8218 $\pm$
.0010 & 12.7 & 32.3 \\
& & &MACHO& $V_{\rm M}$& 816& 39.7 $\pm$ 0.9 & 49\,947.2654 $\pm$
.0009 & 18.1 & 23.3 \\
& & &MACHO &$R_{\rm M}$ & 767& 37.2 $\pm$ 0.9 & 49\,896.1763 $\pm$
.0010 & 17.9 & 32.7 \\
\noalign{\smallskip}\hline\noalign{\smallskip}
V2 & 3.497909 & 0.2858851 & OGLE-II& $I_{\rm C}$ & 501 & 12.4 $\pm$
1.1 & 50\,837.3568 $\pm$ .0041 & 17.5 & 10.4 \\
& & &MACHO& $V_{\rm M}$&1351& 15.2 $\pm$ 0.8 & 49\,938.5438 $\pm$
.0021 & 18.9 & 16.3 \\
& & &MACHO &$R_{\rm M}$ &1188& 13.6 $\pm$ 0.8 & 49\,915.6678 $\pm$
.0027 & 19.9 & 13.0 \\
\noalign{\smallskip}\cline{2-10}\noalign{\smallskip}
& 3.685343 & 0.2713452 & OGLE-II& $I_{\rm C}$ & 501& 4.0 $\pm$ 1.1 &
50\,837.3874 $\pm$ .0118 & 17.5 & 3.3\\
& & &MACHO& $V_{\rm M}$&1351& 3.6 $\pm$ 0.7 & 49\,938.6814 $\pm$
.0089 & 18.9 & 3.9 \\
& & &MACHO &$R_{\rm M}$ &1188& 3.9 $\pm$ 0.8 & 49\,915.6335 $\pm$
.0093 & 19.9 & 3.7 \\
\noalign{\smallskip}\hline\noalign{\smallskip}
V3 & 5.179046 & 0.1930858 & OGLE-II& $I_{\rm C}$ & 471 &  5.1 $\pm$
0.5 & 50\,852.1238 $\pm$ .0027 &  6.8 & 9.3\\
\noalign{\smallskip}\cline{2-10}\noalign{\smallskip}
& 3.495009 & 0.2861223 & OGLE-II& $I_{\rm C}$ & 471& 3.9 $\pm$ 0.4 &
50\,852.0745 $\pm$ .0054 & 6.8 & 7.1 \\
\noalign{\smallskip}\cline{2-10}\noalign{\smallskip}
& 2.005502 & 0.4986283 & OGLE-II& $I_{\rm C}$ & 471& 4.5 $\pm$ 0.5 &
50\,852.2471 $\pm$ .0092 & 6.8 & 8.2 \\
\noalign{\smallskip}\cline{2-10}\noalign{\smallskip}
& 1.684015 & 0.5938189 & OGLE-II& $I_{\rm C}$ & 471& 3.1 $\pm$ 0.4 &
50\,852.1539 $\pm$ .0137 & 6.8 & 5.6 \\
\noalign{\smallskip}\cline{2-10}\noalign{\smallskip}
& 3.816132 & 0.2620454 & OGLE-II& $I_{\rm C}$ & 471& 2.3 $\pm$ 0.5 &
50\,851.9626 $\pm$ .0081 & 6.8 & 4.1 \\
\noalign{\smallskip}
\hline
\noalign{\smallskip}
\end{tabular}

\noindent{\small
$^{\rm a}$ The time is calculated to be as close as possible to the
mean epoch of all observations of a given dataset.  The first two
digits of HJD, that is '24', have been subtracted.}
\end{table*}

\subsection{V3 = OGLE\,051841.98$-$691051.9}
\begin{figure}
\resizebox{\hsize}{!}{\includegraphics{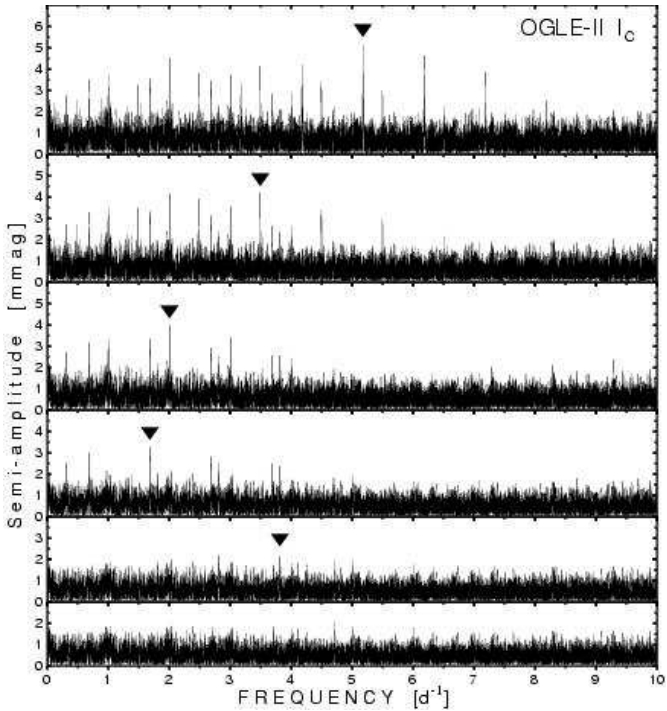}}
\caption{Fourier periodograms of the OGLE-II 1997--2000 $I_{\rm
C}$-filter observations of V3.  From top to bottom the panels show
consecutive steps of prewhitening with frequencies (given in Table 2)
indicated by triangles.}
\label{f_p_v3}
\end{figure}

The third star found to be an LMC $\beta$~Cephei star is about 2~mag
brighter than V1 and V2 (Fig.~\ref{f_cmd}) and shows a very complicated
frequency pattern.  As many as five frequencies with signal-to-noise
(S/N) ratio larger than 4 have been detected (see Sect.~3.5 for the
definition of S/N).  Figure \ref{f_p_v3} shows consecutive steps of
prewhitening for the OGLE-II $I_{\rm C}$ data.  One can
see that all terms have very low amplitudes (Table 2).  The last
detected frequency is rather doubtful.  Moreover, the frequency
2.0055~d$^{-1}$ is very close to 2~(sidereal days)$^{-1}$ which often
appears in periodograms of bright stars.  It is thus probably an
artifact.  However, the other low-frequency term, with $P \approx$
0.5938~d, can be real.  According to the recent calculations of
Pamyatnykh (\cite{pamy99}), massive stars in this region of the H-R
diagram have unstable modes of such long periods, thus the low-frequency 
variation may also represent pulsation.
%
%

Unfortunately, the star was not observed by the MACHO and we
cannot verify the reality of the modes independently, as we did for V1
and V2.  Also, there are not enough datapoints in the OGLE-II $B$
and $V$ band observations to detect reliably a similar variation.

\subsection{Other short-period variables}
As mentioned in Sect.~2, two other
short-period variables may also be $\beta$~Cephei-type stars.  The
first one, V4 = OGLE\,051652.03$-$691405.2 = MACHO\,79.5863.500 is
located in the OGLE-II SC8 and MACHO 79 field.  Its light curve is almost strictly
sinusoidal in shape, has a period $P$ = 0.260083~d, and the
semi-amplitude of about 0.10~mag in all three OGLE-II filters.
There are two characteristics which are difficult to explain if
we assume the star to be a $\beta$~Cephei variable.  First, with $V$
= 17.683, $B-V$ = +0.066, and $V-I_{\rm C}$ = $-$0.008, the star falls
in the colour-magnitude diagram about 1~mag below V1 and V2
(Fig.~\ref{f_cmd}).  Provided that V4 is not $\sim$20~kpc behind the
LMC---which is rather unlikely---this position corresponds to a mid-B-type
star (Fig.~\ref{f_cmd}), too cool to have low-degree $p$ modes
excited.  Second, for such a large amplitude, we would expect to see
non-sinusoidal variations.  For the two $\beta$~Cephei stars with the
largest amplitudes known, BW~Vul and $\sigma$~Sco, the light-curve
shows the so-called 'stillstand', a bump on the rising branch of the light
curve (Sterken et al.~\cite{ster86}; Jerzykiewicz \& Sterken
\cite{jest84}).  Since this is not observed, classifying V4 as a
$\beta$~Cephei variable is open to doubt.

There is, however, an alternative explanation for the variability of
this star.  It could be a very rare contact or nearly contact binary
consisting of two similar mid-B-type stars and seen at low inclination.
Simple calculations indicate that with 2$P$ $\approx$ 0.52~d we can
have a contact binary containing two main-sequence stars with masses
of 5--6~$M_\odot$.  Such an object would have the position of V4 in
Fig.~\ref{f_cmd}.

The next star we suspected to be a $\beta$~Cephei variable, V5 = OGLE
053811.81$-$704245.7 = MACHO 11.9350.116, is very peculiar.  It is found 
in the OGLE-II SC17 and MACHO 11 field.  The star is
fainter than V1 and V2 ($V$ = 17.509), and has much redder
colours ($B-V$ = +0.232, $V-I_{\rm C}$ = +0.351).  However, it
is located very close to the LMC nebula and cluster NGC\,2075 = N\,213
and in addition the SC17 field has a large reddening spread
(see the colour-magnitude diagram presented by Udalski et
al.~\cite{udal00}).  Consequently, its colours could be accounted for
by the reddening effect which will move the star, along the reddening line, 
close to V1 and V2.

\begin{figure}
\resizebox{\hsize}{!}{\includegraphics{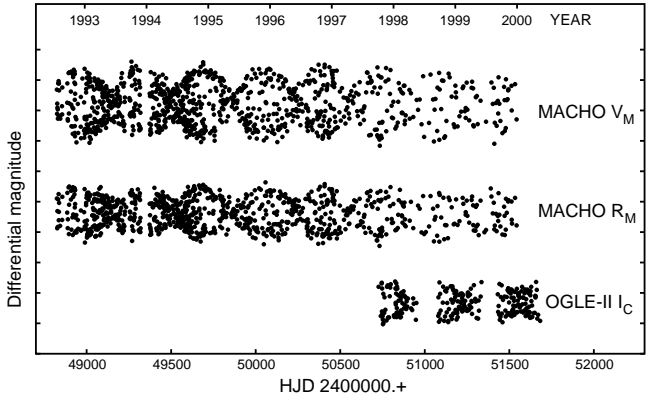}}
\caption{The OGLE-II $I_{\rm C}$ and MACHO $V_{\rm M}$ and $R_{\rm M}$
light curves of V5.  Note the beating due to the presence of two
periodicities with very close frequencies.  Ordinate ticks are
separated by 0.2~mag.}
\label{f_lc_v5}
\end{figure}

If the star were a $\beta$~Cephei variable, it would be indeed an
unusual object.  While the two detected frequencies ($f_1$ =
3.780~d$^{-1}$, $f_2$ = 3.777~d$^{-1}$) are typical for a
$\beta$~Cephei star, the amplitudes are enormous.  The amplitude of the
$f_1$ term in $B$ reaches 0.43~mag, exceeding about twice the
amplitude in BW Vul.  There is also a strong dependence of the amplitude 
on wavelength: in the $I_{\rm C}$ the amplitude is about half that in $B$.  The
second mode has smaller amplitudes, but still rather large for a
$\beta$~Cephei star, 0.24~mag in $B$ and 0.10~mag in $I_{\rm C}$.  The
two frequencies are very close to each other and the beating with a
period longer than 300 days is clearly seen in the data
(Fig.~\ref{f_lc_v5}).  We also found that both periods increase.  The
rate of change of $P_2$ = $f_2^{-1}$ amounts to about 1.7 $\times$
10$^{-8}$ and is seven times larger than that of $P_1$.  As a result of
the changes of the periods, the beat period shortened from 380~d in
1993 to about 317~d in 2000.

%
%

Recalling the arguments given above for V4, we see that almost
sinusoidal variations observed in V5 with even larger amplitudes than
in V4 are difficult to reconcile with $\beta$~Cephei-type variability.
There is, however, another possibility.  The star can be an RR Lyrae
variable of Bailey type c.  The colours, periods, amplitudes and the
shape of the light curve agree well with this type of variability.
Moreover, RRc stars with very close periods have been recently found
(Olech et al.~\cite{olec99a}, \cite{olec99b}; Moskalik \cite{mosk00}).
In the new classification scheme of Alcock et al.~(\cite{alco00}), the
star would be of RR1-$v$1 type.  However, V5 would then be too bright
to be an LMC star.  In accordance with its $V$ magnitude, the star
would be a foreground object located approximately half way to the
LMC.  There is another, indirect argument in favour of this
explanation:  RR Lyraes, especially of the RRc type, are known to
exhibit large erratic period changes (see, e.g., Jurcsik et
al.~\cite{jurc01}, Kopacki \cite{kopa01}) which are rather not
observed in the $\beta$~Cephei stars (Jerzykiewicz \cite{jerz99}).  In
any case, the star is very unusual and deserves further attention.

As far as V5 is concerned, the low-resolution spectroscopy would be
sufficient to distinguish between the two possibilities.  Even
accurate Johnson $U$ photometry, which is probably soon going to
become available (Zaritsky et al.~{\cite{zari97}) would be conclusive
in this case.  For V4 we need radial velocities to decide on the final
classification.

\subsection{Detection threshold}
In view of the results presented here and the discussion that
follows in Sect.~6, it is useful to estimate the detection
threshold of the search.  We therefore analyzed the light curves of
36 variables randomly selected from the
catalogue of \.Zebru\'n et al.~(\cite{zebr01b}) with 14.5 $< V <$ 19.1~mag. 
First, all periodic
signals were removed from the data.  Next, Fourier periodograms of the
residuals were calculated giving mean semi-amplitudes between 0 and 10~d$^{-1}$.
We define this value as the noise (N).  It is easy to find that if we
define signal (S) as the height of the maximum peak in the Fourier
periodogram, the signal-to-noise ratio (S/N) for a pure Gaussian noise
would be (S/N)$_{\rm noise}$ $\approx$ 3.0.  Therefore, we have
arbitrarily set our detection threshold to $D_{\rm th}$ = 4{\rm N}.
The thresholds defined in this way are shown in Fig.~\ref{f_thresh}
for the stars mentioned above.  It can be seen from the figure that
in the range of magnitudes where the LMC $\beta$~Cephei stars are
expected to occur, we should be able to detect amplitudes as low as
3--5~mmag.  This agrees very well with the results given in
Sect.~3.1--3.3 and Table 2.  Since the OGLE-II DIA photometry
practically does not suffer from crowding, the relation shown in
Fig.~\ref{f_thresh} should apply to almost all variables.
%
%
\begin{figure}
\resizebox{\hsize}{!}{\includegraphics{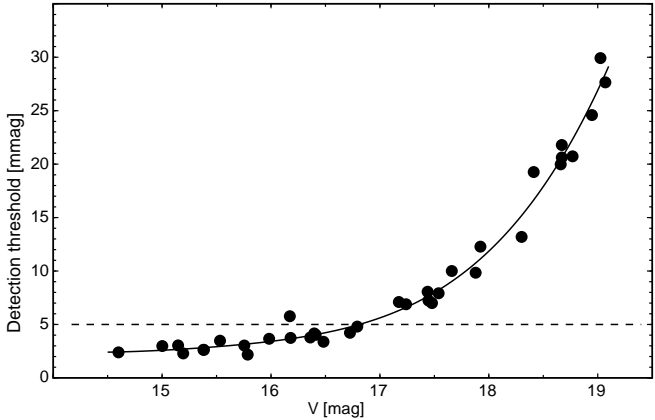}} 
\caption{The
detection threshold, $D_{\rm th}$, for the OGLE-II DIA $I_{\rm C}$-filter
photometry of 36 stars (dots) in the LMC as a function of the $V$ magnitude.  
The continuous line is the polynomial fit to the points, the dashed line shows the
detection threshold of 5~mmag.  See text for the definition of $D_{\rm th}$.}
\label{f_thresh}
\end{figure}

However, we have to remember that the analysis was carried out only on those stars, 
already found as variables by the DIA.  In order to check whether
the DIA missed some small-amplitude variables which could be detected by
periodogram analysis, we calculated S/N for all candidate variables in
the SC1 field using their Fourier periodograms in the range between 0
and 10~d$^{-1}$.  Prior to analysis, we additionally filtered the
observations using a 4$\sigma$ clipping in order to avoid the
influence of outliers.  The S/N values calculated in this way have a
very well defined limit of (S/N) $\approx$ 3.6, independent of the
magnitude.  This shows that the DIA did not miss low-amplitude variables
and its detection threshold is practically the same as that shown in
Fig.~\ref{f_thresh}.

It is also interesting to compare the accuracy of the OGLE-II and
MACHO data.  As can be seen from the comparison of the RSD values in
Table 2, the accuracy of a single observation is slightly better in
the OGLE-II data.  However, because there are usually more
observations for a given star in the MACHO database, the noise levels
in the periodograms are similar for both data sets.

\section{Relation with clusters and associations}

The $\beta$~Cephei-type variability is confined to stars in 
the core-hydrogen burning phase of evolution or shortly beyond this phase having
 masses larger than $\sim$7$M_\odot$
(see, e.g., Pamyatnykh \cite{pamy99}).  From the comparison with evolutionary 
models it can easily be shown that a $\beta$~Cephei star cannot be older than
$\sim$35~Myr.  This is why almost all Galactic $\beta$~Cephei stars can be
identified as the members of the young open clusters or OB associations. 
We therefore expect that
the LMC $\beta$~Cephei-type stars can also be members of clusters or associations.

\subsection{V1 and the LH\,81 association}
Our first variable, V1, is located in the middle of the ring-shaped
superbubble N\,154 (Henize \cite{heni56}) = DEM\,246 (Davies et
al.~\cite{davi76}) which is also a diffuse X-ray source (Dunne et
al.~\cite{dunn01}).  The two brightest parts of N\,154, south-western
BSDL\,2434 (Bica et al.~\cite{bica99}) and the much brighter
north-eastern LH\,87n, with a bright knot NGC\,2048, are well seen in
Fig.~\ref{f_assoc}.  N\,154 encompasses two large OB associations:
LH\,81 (Lucke \& Hodge \cite{luho70}) = SL\,589 (Shapley \& Lindsay
\cite{shli63}) and LH\,87 (see Fig.~\ref{f_assoc}).  Several compact
groups within LH\,81 are usually distinguished as open clusters; the
brightest one is the OGLE-CL-LMC605 (Pietrzy\'nski et al.~\cite{piet99}) =
BCDSP\,8 (Bica et al.~\cite{bica96}).  The LH\,81 association has been
studied by Massey et al.~(\cite{mass00}) and is known to contain many
very hot and massive objects including three Wolf-Rayet stars and two
stars of spectral type O3-4.  Figure \ref{f_cmdass} shows the
colour-magnitude diagram for a region in LH\,81 with a radius of
2${\farcm}$7 centered at ($\alpha, \delta$)$_{\rm 2000.0}$ = (5$^{\rm
h}$34$^{\rm m}$39$^{\rm s}$, $-$69$\degr$43$\arcmin$11$\arcsec$).  We
see that the main sequence extends to stars as bright as eleventh
magnitude and that V1 is a likely member of the association.  If this
is true and the star formation was coeval, V1 should not be older than
5~Myr, the age of LH\,81 deduced by Massey et al.~(\cite{mass00}).
%
%
\begin{figure*}
\resizebox{\hsize}{!}{\includegraphics{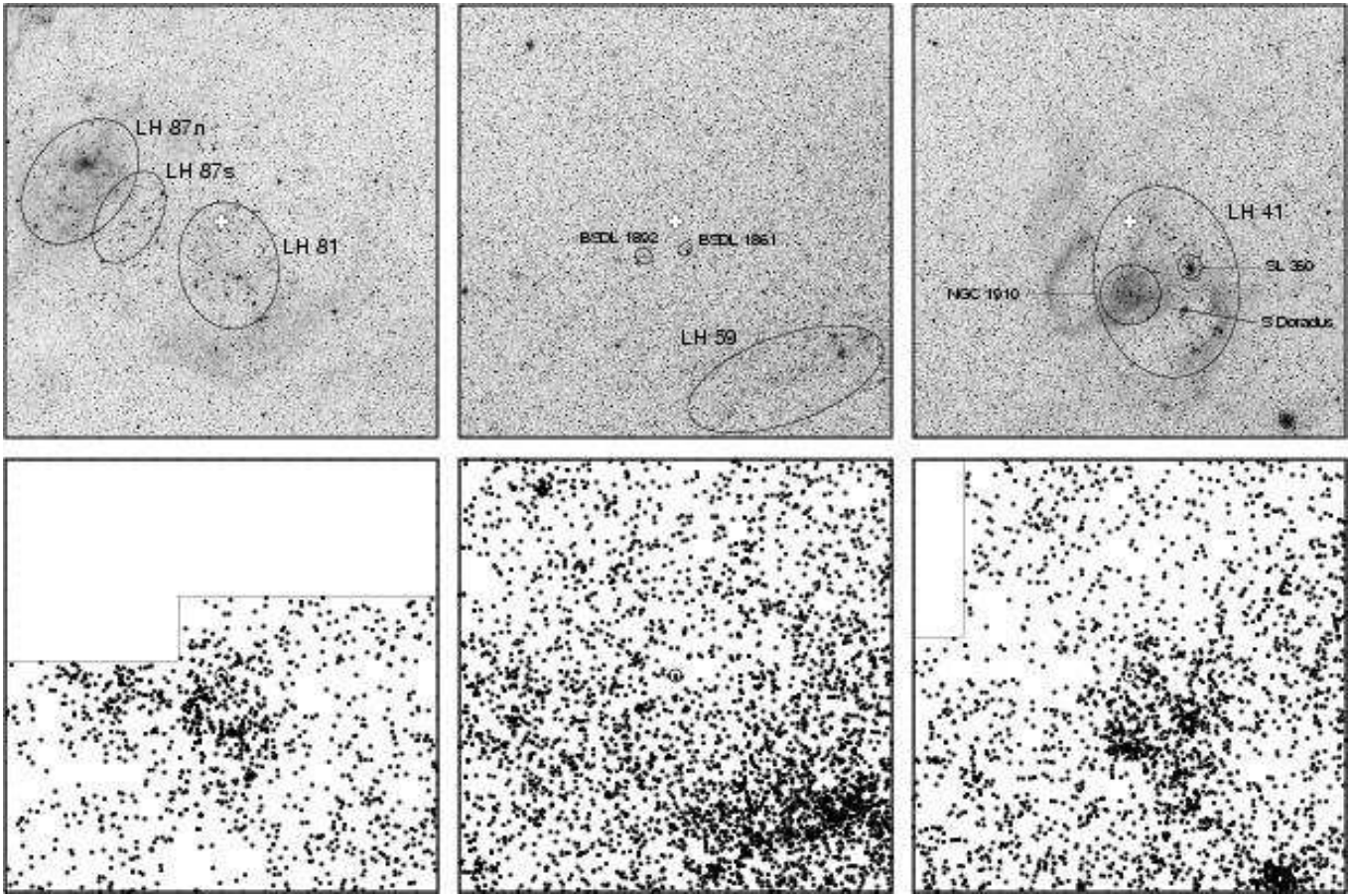}}
\caption{Three 20$\arcmin$ $\times$ 20$\arcmin$ fragments of the
red-filter POSS-II plates centered at V1 (left), V2 (middle), and V3
(right).  Associations, clusters and stars discussed in the text
are labeled.  White crosses mark the positions of $\beta$~Cephei
variables.  Below, for each variable (encircled) the location of
nearby bright, hot stars with $V <$ 18~mag and $(V-I_{\rm C}) <$
0.5~mag, taken from the catalogue of Udalski et al.~(\cite{udal00}), is
shown on the same scale.  North is up, east to the left.}
\label{f_assoc}
\end{figure*}
%
%
\begin{figure*}
\resizebox{\hsize}{!}{\includegraphics{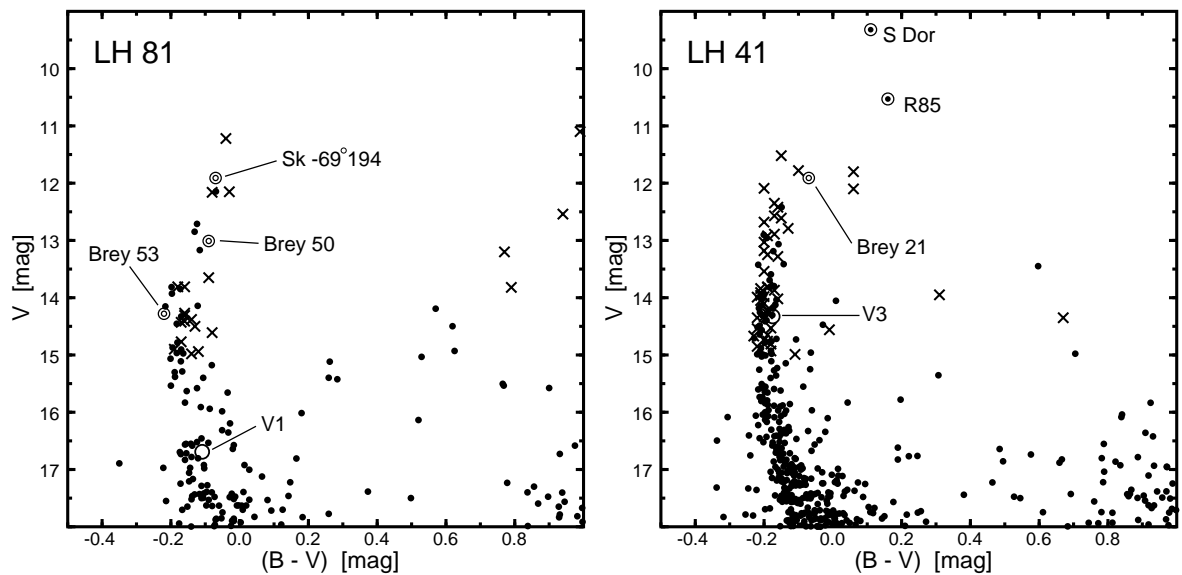}}
\caption{Colour-magnitude diagrams for the brightest stars of the
LH\,81 (left) and LH\,41 (right) association.  The photometry of stars
shown as dots was taken from Udalski et al.~(\cite{udal00}), those
shown as crosses, from Massey et al.~(\cite{mass00}).  The Wolf-Rayet
stars are shown as encircled open circles, LBVs, as encircled dots and
$\beta$~Cephei stars, as open circles, in addition to a label.   Only
stars with $(B-V) <$ 1.0~mag and $V <$~18~mag are shown in the plots.}
\label{f_cmdass}
\end{figure*}

\subsection{V2 and the LH\,59 association}
V2 is located about 7$^\prime$ north-east from the center of the LH\,59
association (Fig.~\ref{f_assoc}).  The nearest clusters are
the OGLE-CL-LMC499 = BSDL\,1861 and OGLE-CL-LMC505 = BSDL\,1892.  The colour-magnitude
diagrams of the clusters provided by Pietrzy\'nski et
al.~(\cite{piet99}) show that at least BSDL\,1892 contains some hot
stars.  The ages of the two clusters estimated by Pietrzy\'nski \&
Udalski (\cite{piud00}) are 80 and 250~Myr for BSDL\,1861 and 1892,
respectively.  Thus, the clusters are much too old to contain
$\beta$~Cephei stars.  However, LH\,59 seems to be slightly younger.
Dieball \& Grebel (\cite{digr00}) estimated the age of the three rich
clusters embedded in this association (NGC\,1969, 1971, 1972) to be in
the range 40--70~Myr.  Taking into account the errors in the age estimates
and the possibility that the formation was not strictly coeval in such
a large complex, the presence of a $\beta$~Cephei star in this region
might still be possible.  Although we cannot be certain that V2 belongs to
LH\,59, the density of hot stars in this region is quite high.  We
conclude that V2 belongs either to the coronal region of LH\,59 or to
the LMC field, where, as found e.g.~by Massey et al.~(\cite{mass95})
and Holtzman et al.~(\cite{holt99}), the star formation is still in
progress.

We also note that V2 has been detected as a far-ultraviolet source by
the UIT telescope.  Its B1-filter magnitude of 14.07 $\pm$ 0.07
(Parker et al.~\cite{park98}) indicate that it is really an early
B-type star.

\subsection{V3 and the LH\,41 association}
It is, in turn, easy to indicate a connection of our third variable,
V3, with an LMC association.  The star is located in the rich LH\,41
association known to contain S\,Doradus, the extreme Luminous Blue
Variable (LBV, Humphreys \& Davidson \cite{huda94}).  The association
(Fig.~\ref{f_assoc}) contains also another probable LBV star, R\,85, a Wolf-Rayet star,
Brey 21, and many O-type stars (Breysacher et al.~\cite{brey99};
Massey et al.~\cite{mass00}).  Several young clusters including the compact
SL\,360 = HD\,35342 and the larger NGC\,1910 = SL\,371, form the
well-seen groups within the association.  The whole region is
surrounded by the N\,119 = DEM\,132 nebula (Fig.~\ref{f_assoc}).
There is no doubt that intensive star formation is taking place here.
The age of the association estimated by Massey et al.~(\cite{mass00}) is
4~Myr.  The colour-magnitude diagram for the region of LH\,41 with a
4$\farcm$5 radius centered on ($\alpha, \delta$)$_{\rm 2000.0}$ =
(5$^{\rm h}$18$^{\rm m}$30$^{\rm s}$,
$-$69$\degr$13$\arcmin$15$\arcsec$) is shown in Fig.~\ref{f_cmdass}.

The $UBV$ photometry of V3 has been obtained by Massey et
al.~(\cite{mass00}).  The star has been designated as LH41-64 by these
authors and has $V$ = 14.32, $B-V$ = $-$0.16 and $U-B$ = $-$1.02~mag. Thus
we get $E(B-V)$ = 0.14~mag and
intrinsic colours:  $(B-V)_0$ = $-$0.30 and $(U-B)_0$ = $-$1.12~mag
which correspond rougly to a B0/O9 type star.

\section{Notes on previous searches}
As mentioned in the Introduction, the previous searches for
$\beta$~Cephei stars in the LMC (Sterken \& Jerzykiewicz \cite{stje88};
Kubiak \cite{kubi90}; Balona \cite{balo93})
yielded some variables,
although none of them could be convincingly classified as a $\beta$
Cephei variable.  With the OGLE-II and MACHO photometry at hand we can now
try to verify their variability.

Out of the six program and five comparison stars from Brunet et
al.~(\cite{brun75}) selected for observations by Sterken \&
Jerzykiewicz (\cite{stje88}), only star 90 and 134 are within
the OGLE-II fields.  Both stars were identified by the DIA as variables.  However, the
amplitudes are very small (less than 0.006~mag for star 90 and
0.01~mag for star 134) and are very likely of instrumental origin.  On
the other hand, the MACHO observed all these stars but 230.  From our
analysis of the ten stars we found that:  (i) none of the stars shows
short-period variability, (ii) stars 32 and 257 are variable, but the
light changes are aperiodic and have the amplitudes below 0.2~mag, (iii) a
1-year periodicity is seen for four stars (90, 146, 230, and 134), but 
comparing the OGLE-II photometry of
stars 90 and 134 we are certain that the variations are of instrumental
origin, (iv) stars 269 and 271 are heavily blended.  Concluding, none
of the stars show $\beta$~Cephei-type variability.

%
\begin{table}
\caption{Stars found to be variable in previous searches for
$\beta$~Cephei stars in the LMC identified with MACHO sources}
\begin{tabular}{rll}
\hline\noalign{\smallskip}
Star & MACHO name & Comment on variability \\
\noalign{\smallskip}\hline\noalign{\smallskip}
\multicolumn{3}{c}{Sterken \& Jerzykiewicz (\cite{stje88})}\\
\noalign{\smallskip}\hline\noalign{\smallskip}
   32 & 18.2479.10 & Irregular variations\\
   90 & 79.4774.10 & Constant \\
   146& 49.6736.14 & Constant \\
   210& ~~8.8784.291 & Constant \\
   257& 50.9639.7 & Irregular variations \\
\noalign{\smallskip}\hline\noalign{\smallskip}
\multicolumn{3}{c}{Kubiak (\cite{kubi90})}\\
\noalign{\smallskip}\hline\noalign{\smallskip}
    1 & 44.1626.20 & Constant \\
    2 & 44.1626.35 & Long-period variable \\
    3 & 44.1626.46 & Eclipsing, P = 2.291265 d \\
    4 & 44.1626.46 & Eclipsing, P = 1.636929 d \\
    7 & 44.1747.28 & P = 0.9403 d ?\\
    9 & 44.1868.9 & Irregular \\
\noalign{\smallskip}\hline\noalign{\smallskip}
\multicolumn{3}{c}{Balona (\cite{balo93})}\\
\noalign{\smallskip}\hline\noalign{\smallskip}
   87 & 61.8192.120& Eclipsing, P = 1.662846 d\\
  241 & 61.8192.154& Irregular variations\\
  297 & 61.8192.55 & Constant \\
  491 & 61.8191.12 & Constant \\
  682 & 61.8191.81 & Constant \\
\noalign{\smallskip}\hline
\end{tabular}
\end{table}

Kubiak (\cite{kubi90}) found nine variables in the LMC cluster
NGC\,1712.  Of these, seven (all but 5 and 8) are included in the MACHO
field 44.  Variable 6 is a blend and we could not cross-identify it properly,
but the three nearby MACHO stars are constant.  Of the remaining six
stars, only variable 1 seems to be constant.  Variable 2 shows
a smooth increase and then decrease of brightness with an amplitude of
0.4~mag in $V_{\rm M}$ and 0.5~mag in $R_{\rm M}$.  Variables 3 and 4
are eclipsing binaries with Algol-type light curves.  The eclipses of equal depth
(0.42~mag for variable 3 and 0.17~mag for variable 4) indicate that
the binaries have equal components.  Variable 7 has a small amplitude
and a probable period of 0.94~d, while variable 9 shows irregular
variations mainly in the $V_{\rm M}$ band.

Balona (\cite{balo93}) detected variability of seven stars in
another LMC cluster, NGC\,2004.  Of those, five have photometry in the
MACHO database.  We confirm the variability of only two (Table 3).
Star 87 is an eclipsing binary with a $\beta$~Lyrae-type light curve and
an orbital period of 1.662846~d while
star 241 is an irregular variable.  The period of 1.835~d given by
Balona (\cite{balo93}) for the latter star is not confirmed.  Stars
297, 491, and 682 were found to be constant.

The other LMC cluster searched by Balona (\cite{balo93}), NGC\,2100, was 
not observed by either the MACHO or the OGLE-II.

\section{Discussion}

$\beta$ Cephei-type stars
have been finally found in the LMC. These are the first extragalactic 
stars of this type so far discovered. Photometric indices,
magnitudes, amplitudes and periods of three variables 
fit the $\beta$~Cephei characteristics.  However,
the spectroscopic confirmation and/or the spectral type of these
variables would be, of course, desirable.   

Since there are over 27,000 early B-type stars in the LMC area
covered by the OGLE-II observations, and the variability detection threshold
was rather low, we may conclude that for the LMC the BCIS has been mapped.  
Only three $\beta$~Cephei-type variables have been found.  Two important facts arise
from this finding.  The first is the fraction of $\beta$~Cephei stars
with respect to the whole population of early B-type stars, the other is their
position in the colour-magnitude diagram.

\subsection{Fraction of $\beta$~Cephei stars with respect to early
B-type stars in the LMC}

Practically all
Galactic $\beta$~Cephei stars have $M_{\rm V}$ between $-$4.5 and
$-$1.5~mag.  We take this range as a reference and in
order to avoid selection effects we restrict the comparison to
stars in open clusters.  Moreover, to make
the comparison reliable, we need to account for different detection
thresholds.  A detailed analysis of
the fraction of $\beta$~Cephei stars in Galactic open
clusters is going to be published soon (Pigulski et al., in
preparation), here we mention only the preliminary result presented by
Pigulski et al.~(\cite{pigu02}).  These authors pointed out that the
fraction of $\beta$~Cephei stars in Galactic southern clusters is few
times larger than in the northern ones, and, if related to all
main-sequence stars falling into the $M_{\rm V}$ range given above 
it amounts to 35 $\pm$ 7\% for the southern and 5.3 $\pm$ 1.7\% for the
northern clusters.  Since southern clusters are closer to the
Galactic centre, this was explained as being a consequence of the Galactic
metallicity gradient.  At the LMC distance, $-$4.5 $\leq M_{\rm V} \leq$
$-$1.5~mag corresponds roughly to 14.3 $\leq V \leq$ 17.3~mag.
There are 27,663 main-sequence stars in the OGLE-II fields falling in this
range of $V$ magnitude.  (As a main-sequence star we mean here a star with $(V -
I_{\rm C}) <$ 0.5~mag; since contamination by foreground stars is
very low in this part of the colour-magnitude diagram, it does not
affect our estimate.) The three $\beta$~Cephei stars we found,
constitute 3/27663 $\approx$ 0.011 $\pm$ 0.006\% of early B-type stars
(we assume a Poisson statistics to derive errors).  Even if we take
into account the fact that for searches in the Galactic open clusters
roughly twice as low threshold is achieved as that seen in
Fig.~\ref{f_thresh}, so that some small-amplitude variables could have
been missed in the LMC, the striking difference in the $\beta$~Cephei
fraction between the Galaxy and the LMC is evident.  In other words, as
the average LMC metallicity is [M/H] $\approx$ $-$0.4 (or $Z$
$\approx$ 0.007), this is another observational confirmation of the
fact that the driving mechanism strongly depends on the metallicity.

\subsection{Position of $\beta$~Cephei-type stars in the colour-magnitude 
diagram: a comparison
with theoretical predictions for LMC metallicity}
Should we, however, expect to observe $\beta$~Cephei stars at all in the LMC if
$Z$ = 0.007?  The two recent theoretical predictions (Pamyatnykh
\cite{pamy99}; Deng \& Xiong \cite{dexi01}) give different answers to
this question.  Pamyatnykh (\cite{pamy99}, see his Fig.~11) for $Z$ =
0.01 does not find unstable modes for stars with masses smaller than
25~$M_\odot$.  On the other hand, Deng \& Xiong (\cite{dexi01})
predict instability for stars with masses $\sim$10$M_\odot$ down to
$Z$ = 0.005.  The following reasons coming from our LMC results 
favour the predictions of Pamyatnykh (\cite{pamy99}):
\begin{itemize}
\item First, we find $\beta$~Cephei stars both among low mass (8--10
$M_\odot$, V1 and V2) and high mass (25--30~$M_\odot$, V3) stars.  In
Fig.~11 of Pamyatnykh (\cite{pamy99}), the BCIS is shown for three
values of $Z$:  0.01, 0.015 and 0.02.  We see in this figure that for
$Z$ = 0.015 the lower part of the BCIS covers about half of the
main-sequence width.  Going towards higher masses the BCIS becomes
narrower and then widens again.  For $Z$ = 0.01 there is no lower part
at all and instability begins for masses $\sim$25~$M_\odot$.  It seems
that for a value of $Z$ between 0.01 and 0.015 the BCIS splits in two regions,
one for masses around 10~$M_\odot$
and one over 25~$M_\odot$.  This does correspond to the
situation observed in the LMC.
\item In order to obtain the agreement mentioned above, we need,
however, to have either a higher metallicity in the LMC $\beta$~Cephei stars or 
the splitting of the BCIS described above, but for $Z$ = 0.007.
Although Pamyatnykh (\cite{pamy99}),
comparing the location of the hot border of the BCIS based on OP and OPAL
opacities, estimates that an error in $Z$ of about 0.003--0.005 can be
involved, we think that the first possibility is more probable.  We
know from Sect.~4 that at least two, and perhaps all three
$\beta$~Cephei stars in the LMC are located in regions of violent star
formation.  The recent studies of the age-metallicity relation for the LMC
(Cohen \cite{cohe82}; Pagel \& Tautvai{\v s}ien\.e \cite{pata98}) and
direct metallicity determinations in young objects (Olszewski et
al.~\cite{olsz91}; Luck \& Lambert \cite{lula92}; Jasniewicz \&
Th\'evenin \cite{jath94}; Rolleston et al.~\cite{roll96}; Luck et
al.~\cite{luck98}; Korn et al.~\cite{korn00}) indicate that although
the mean LMC metalicity is about [M/H] $\approx$ $-$0.4, a large
spread in metallicities of young objects is observed.  It is even
possible that some objects have solar or nearly solar metallicity.  In
fact, it would be na\"{\i}ve to believe that dozen megayears of the LMC
history did not leave any metallicity inhomogeinities.  It is
therefore very likely that our variables have a metallicity higher than
the LMC average.  Nevertheless, spectroscopic determination of
abundances in these stars would be highly desirable.
\item Why do we think the predictions of Deng \& Xiong (\cite{dexi01})
are incorrect?  First, these authors do not find the instability among
massive stars, so it would be difficult to explain the variability of
V3 with a lower than Galactic metallicity.  Next, with instabilities
predicted down to $Z$ = 0.005, we should have found many more $\beta$~Cephei
stars than we did.  Finally, Deng \& Xiong (\cite{dexi01}) used
the first 1992 version of the OPAL opacity tables with an analytical
approximation which---as these authors write---gives as much as 10\%
differences with respect to the tabulated values.  In view of the
subtle domination of driving over damping, this could influence
considerably the final result.
\end{itemize}

\section{Summary and future work}
The most important results obtained in this paper can be summarized
as follows:
\begin{enumerate}
\item Three $\beta$~Cephei stars have been found in the LMC bar.
\item Two of them were found to have $V$ magnitude corresponding
roughly to the LMC B2--B2.5 stars, while the third one (V3) is as bright
as a B0-type star.
\item The three variables constitute a very small fraction (less than
0.02\%) of all early B-type stars falling within the range of
magnitudes where the Galactic stars of this type are distributed.  This is
much less than for the Galactic clusters and can be explained by a strong
dependence of the driving mechanism on metallicity.
\item The location and number of $\beta$~Cephei variables in the LMC is in
reasonable agreement with the theoretical predictions of Pamyatnykh
(\cite{pamy99}) provided that the metallicity of these stars is
slightly higher than the LMC average for young population, namely,
between $Z$ = 0.01 and 0.015 ([M/H] between $-$0.3 and $-$0.1).
\item All three variables were found to be in or very close to regions
of intensive star formation, implying that they indeed can have
enhanced metallicities.
\item If the low-frequency variation in V3 is confirmed, this can be
an interesting example of a luminous $\beta$~Cephei star in which
modes with longer than typical periods are excited.  The star could be
therefore a good target for asteroseismology.
\end{enumerate}

The main goal of this series of papers is to give a thorough
observational picture of the variability among the LMC and SMC stars lying
in the upper main sequence, where the $\kappa$ mechanism related to
the metallicity bump is at work.  For this purpose we shall use, like
in the present paper, the OGLE-II and MACHO photometric databases.  In
particular, we are going to focus on the study of slowly pulsating B
stars (SPB), which, although fainter than $\beta$~Cephei variables,
have typical amplitudes large enough to be detected in the LMC and even in
the SMC.  Next, we are going to search bright B and A-type stars for
variability with periods of the order of 1 day and longer.  This would
probably shed some light on the presence of the $g$ modes in luminous
early-type stars.  Finally, the variability of Be stars and
hot eclipsing binaries will be studied as well.

\begin{acknowledgements}
The paper has been supported by the KBN grant No.~2\,P03D\,006\,19.
We thank Prof.~M.\,Jerzykiewicz, Drs.~G.\,Kopacki and A.\,Pamyatnykh
for fruitful discussions.  The detailed comments of the anonymous
referee are acknowledged. We also thank the staff of the Institute of
Mathematics of our University, especially Mrs.~M.~Ko{\l }aczkowska,
for making available their computing facilities.

This paper utilizes public domain data originally obtained by the
MACHO Project, whose work was performed under the joint auspices of
the U.S.~Department of Energy, National Nuclear Security
Administration by the University of California, Lawrence Livermore
National Laboratory under contract No.~W-7405-Eng-48, the National
Science Foundation through the Center for Particle Astrophysics of the
University of California under cooperative agreement AST-8809616, and
the Mount Stromlo and Siding Spring Observatory, part of the
Australian National University.
\end{acknowledgements}

\end{document}